\documentstyle[aps,multicol,epsf]{revtex}
\begin{document}
\draft
\title{ A Generalized Ginzburg-Landau Approach to   
Second Harmonic Generation }
\author{Debanand Sa\cite{e-dsa}}
\address{Institut f\"ur Physik, Universit\"at Dortmund,  
44221 Dortmund, Germany}  
\author{R. Valent\'\i\ and C. Gros }
\address{Fachbereich Physik, University of the Saarland,  
66041 Saarbr\"ucken, Germany}  
\maketitle

\begin{abstract}
We develop a generalized Ginzburg-Landau theory for second 
harmonic generation (SHG) in magnets
 by expanding the free energy in terms of  
the order parameter in the magnetic phase and the
 susceptibility tensor in the
 corresponding high-temperature phase. The non-zero components
of the SHG susceptibility in the ordered phase are derived from
  the symmetries of the  susceptibility tensor in the high-temperature
 phase  
and the symmetry of the order parameter.
 In this derivation, the dependence of the SHG susceptibility
 on the order parameter follows naturally, and therefore its 
nonreciprocal optical properties.  
 We examine this phenomenology for the magnetoelectric compound  
Cr$_2$O$_3$ as well as for the ferroelectromagnet  
YMnO$_3$.  
\end{abstract}

\begin{multicols}{2}
\narrowtext
Second Harmonic Generation (SHG) is a very useful technique to study the  
nonlinear optical properties \cite{sh} in magnetic materials. The recent 
observation \cite{kri,fie} of nonreciprocal optical effects (i.e., not  
invariant under time reversal operation) in the magnetoelectric material 
Cr$_2$O$_3$ below the N\'eel temperature $T_N$ has been of great 
importance in 
the study of antiferromagnetic (AFM) ordering by light. With the
 help of SHG, photographs of the antiferromagnetic domains
 \cite{fie,fie1} in
 Cr$_2$O$_3$ were obtained what confirms that these experiments
can distinguish between the two magnetic states that are related
 to each other by the time reversal operation, and therefore
 its nonreciprocity.  These observations
 can be explained by an interference effect between a time-symmetric
 magnetic dipole contribution and a time-antisymmetric electric
 dipole contribution \cite{fie}. Soon after these experiments were done,
 a microscopic theory was proposed \cite{mut} which could 
 explain quantitatively the non-reciprocal effects observed
 in Cr$_2$O$_3$. It was shown that the electric dipole contributions
 were linearly proportional to the antiferromagnetic order parameter
 giving rise to the time-antisymmetric character to the electric dipole
 tensor. The study of this dependence is of  crucial
importance in order to understand the non-reciprocal character
 of the SHG tensors. The derivation of a micoscopic theory for
 a specific effect, in our case SHG, can
 be rather complicated depending on the properties of the material
 under study \cite{mut,tanabe}.
 Concrete information of the SHG process in materials where
 a transition takes place, i.e. paramagnet-antiferromagnet or 
 paraelectric- ferroelectric, can be obtained
at a simpler level, i.e. by considering only the symmetry arguments.
  In particular, we are interested in investigating the dependence
 of the SHG susceptibilities in the ordered phase on the order
 parameter. 

 A powerful phenomenological theory suitable
 to describe phase transitions is the Ginzburg-Landau theory.
   The Ginzburg-Landau approach is based on the existence of an
 order parameter in the ordered phase and 
on symmetry considerations \cite{ginzburg}.
    Pershan in 1963 \cite{per} showed that the tensors for 
nonlinear electro- and magneto-optic effects could be 
 derived from a phenomenological
 time-averaged free energy.  It is our purpose in this paper
 to extend the formulation of Pershan by including the order
 parameter explicitly in the 
 expansion of the free energy for SHG.  We shall show
 that the non-zero components of the SHG susceptibility
 tensor in the ordered phase are naturally obtained
 from the symmetry of the susceptibility tensor in the
 high-temperature phase and the symmetry of the
 order parameter. Once the dependence of the SHG susceptibility
tensor
 on the order parameter is known, the non-reciprocal optical 
 properties below the transition temperature follow.     
 We explicitly 
verify this formulation in the magnetoelectric compound Cr$_2$O$_3$ as 
well as the hexagonal ferroelectric-antiferromagnetic material YMnO$_3$. 

A convenient starting point in order to describe
the optical nonlinearities is given by a
time-averaged free energy $F$ as proposed by Pershan
\cite{per}. A dipole expansion of the
induced current 
${\bf J}=\partial{\bf P}/\partial t + c\nabla\times{\bf M}
-\partial(\nabla\cdot{\bf Q})/\partial t+\dots$,
where
{\bf P}, {\bf M} and {\bf Q} are respectively the 
electric dipole polarization, the magnetization
and the electric quadrupole polarization, leads to \cite{per}

\begin{equation}  
F = - 2Re\sum_{\nu=1}^{n}\left[\, 
{\bf {E^\ast}}(\omega_{\nu})\cdot {\bf P}(\omega_ 
{\nu})+ {\bf {H^\ast}}(\omega_{\nu}) \cdot {\bf M}(\omega_{\nu}) 
\,\right]~,
\label{F} 
\end{equation} 
where we have discarded the electric quadrupole and higher-order
terms and a constant contribution. Here  
${\bf E}(\omega_\nu)$,  ${\bf P}(\omega_\nu)$, 
${\bf H}(\omega_\nu)$ and ${\bf M}(\omega_\nu)$ are respectively the 
Fourier components of the electric field, electric dipole polarization, 
magnetic field and the magnetization  and
$\nu$ denotes the number of partial waves. The usual relations

\begin{eqnarray}  
{\bf P}(\omega_{\nu}) =  - \partial{F} / \partial{\bf {E^\ast}}(\omega_{\nu})  
\nonumber \\
{\bf M}(\omega_{\nu}) =  - \partial{F} / \partial{\bf {H^\ast}}(\omega_{\nu})  
\nonumber   
\end{eqnarray}  
are fullfilled. The electric dipole polarization and the 
magnetization are, in general,
 nonlinear functions of {\bf E} (and {\bf H})
 which may be expanded into a power
series in {\bf E} (and {\bf H}). 
The second order term for the electric dipole polarization is

\begin{eqnarray} 
P_i (2\omega) =
\chi_{ijk}^{(ED)}(2\omega,\omega,\omega) \> E_j(\omega) \> E_k(\omega)
\end{eqnarray} 
which corresponds to the SHG electric dipole contribution (ED).
The corresponding free energy is then (see Eq.\ (\ref{F})):
   
\begin{eqnarray}  
F^{(ED)} = -\left[\,\chi_{ijk}^{(ED)}(2\omega,\omega,\omega) 
\> E_i^\ast (2\omega) \> E_j(\omega) \> E_k(\omega) 
\nonumber \right.\\ \left.  
+{\chi^{*}_{ijk}}^{(ED)}(2\omega,\omega,\omega)
\> E_i (2\omega) \> E_j^\ast (\omega) \> E_k^\ast (\omega)
\,\right] \label{F_ED}
\end{eqnarray} 
 Summation over repeated
indices is implicit in the above formulas.

As a second example, we consider the magnetic
dipole contribution (MD) to SHG,

\begin{eqnarray} 
M_i (2\omega) =
\chi_{ijk}^{(MD)} (2\omega,\omega,\omega)\> E_j(\omega) \> E_k(\omega)~,
\end{eqnarray}  
leading to

\begin{eqnarray}
F^{(MD)} = -\left[\,\chi_{ijk}^{(MD)}(2\omega,\omega,\omega) 
\> H_i^\ast (2\omega) \> E_j (\omega) \> E_k (\omega)  
\nonumber \right.\\ \left. 
+{\chi^{*}_{ijk}}^{(MD)}(2\omega,\omega,\omega)
\> H_i (2\omega) \> E_j^\ast (\omega) \> E_k^\ast (\omega)
\,\right]~. \label{F_MD}  
\end{eqnarray}
$F^{(MD)}$ is the free energy corresponding to the
MD contribution to SHG.
Both the ED and the MD
contribute to the source term S, i.e., as 
$\mu_0 [(\partial^{2}{P}/ \partial{t^2})    
+ \nabla \times (\partial{M}/ \partial{t})]$ in the wave     
equation for the electric field, derived from the Maxwell's 
equations in classical electrodynamics \cite{fie}.
The measured output intensity in a nonlinear optical
experiment is 
$I \propto {\mid S \mid}^2$ and once the
 SHG susceptibility in the ordered 
phase is known, 
possible non-reciprocal properties of the system  may be derived.
 
In the context of the antiferromagnetic Cr$_2$O$_3$ it has been 
observed \cite{fie} that $\chi_{ijk}^{(ED)}$ exists only in the
ordered phase ($T<T_N$) and it has been shown \cite{mut,tanabe}
 that it
is linearly 
proportional to the antiferromagnetic order parameter, as it should
 be if we had defined a
 Ginzburg-Landau free energy for the ordered phase.
In the framework of the classical Ginzburg-Landau approach
to phase transitions the allowed terms contributing
 to the free energy follow from the
symmetry of the order-parameter and from
the symmetry of the lattice \cite{ginzburg,tol}.
It is therefore of interest to ask oneself whether one can 
combine the standard Ginzburg-Landau approach for the
magnetic properties of e.g.\ Cr$_2$O$_3$ with the
expressions Eq.\ (\ref{F_ED}) and Eq.\ (\ref{F_MD})
in order to obtain a more general formulation for the nonlinear
magneto-optical properties of a given compound.

In order to generalize the Ginzburg-Landau formulation 
to study the nonlinear optical properties in antiferromagnets 
we first note that the (generalized) Ginzburg-Landau functional
has to obey the symmetry of the high-temperature phase, 
as the (spontaneous) breaking of this symmetry is inherent in
the solution which is a minimum of the Ginzburg-Landau
functional. As an illustration,  
we consider a hypothetical magnetic dipole contribution to SHG
 which is only present in the ordered phase.  Let us assume 
 the following symmetries for the order parameter. i) it is
 a c-tensor, i.e. antisymmetric under the time reversal operation
 and ii) it is a (real) {\it {axial}} tensor of first rank
( a pseudovector),
  i.e.\ $O_l$. In this case
  
\begin{eqnarray}
F = -\left[\,\chi_{ijkl}(T>T_N) \> H_i^\ast \> E_j \> E_k \> \right.   
\nonumber \\ \left.   
+\ \chi^\ast_{ijkl}(T>T_N) \> H_i \> E_j^\ast \> E_k^\ast
\,\right] \> O_l  \label{example}
\end{eqnarray}
 would be a valid expression for the combined free-energy,
where $\chi_{ijkl}(T>T_N)$ is the susceptibility tensor in the 
paramagnetic phase. The expression in the parenthesis indicates that the  
susceptibility is  above the transition
 temperature $T_N$. 
It should be noted here that the free energy is  
a i-scalar, i.e., a scalar invariant under the 
time reversal operation. This 
implies that the susceptibility tensor in the paramagnetic  
phase has to be a polar i-tensor of fourth rank.  
Now, comparing Eq.\ (\ref{example}) with Eq.\ (\ref{F_MD}), one
would obtain    
the following relation between the SHG susceptibility and 
the order parameter: 

\begin{equation}  
\chi_{ijk}^{SHG}(T<T_N) = \chi_{ijkl}(T>T_N)\> O_{l}  
\end{equation}  

 It is clear from the above expression that from the knowledge 
of the order parameter and the symmetries of the susceptibility tensor 
\cite{bir} in the paramagnetic phase, it is possible to obtain all the   
non-zero components of the SHG susceptibility in the ordered phase.
 Moreover, the SHG 
susceptibility tensor below the transition temperature becomes directly 
proportional to the order parameter, which will ultimately manifest in 
 the non-reciprocal optical effects in the system. 
In what follows, we examine 
this in detail for the cases of the 
magnetoelectric compound Cr$_2$O$_3$ and of the 
ferroelectric-antiferromagnetic material YMnO$_3$.  

Cr$_2$O$_3$, in its paramagnetic phase (above $T_N\approx 307$ K), 
crystallizes in the centrosymmetric point group $\overline{3}m$. The unit  
cell contains four Cr$^{3+}$ ions, which occupy equivalent c-positions  
along the optical axis. This structure has a centre of inversion 
and parity 
considerations allow only axial i-tensors of odd rank and polar i-tensors 
of even rank. Thus, above $T_N$, SHG electric dipole effects are 
forbidden but magnetic dipole effects are allowed. Below $T_N$, the
four spins 
in the unit cell order along the optical axis in a non-centrosymmetric AFM 
structure (the spin order being, up, down, up, down) which leads to two 
types of domains transformed into each other 
by time reversal symmetry. Both, space and time reversal symmetry operations 
are separately broken by the spin ordering
 but the combination of both the symmetries  remains a symmetry of the 
crystal. 
The magnetic point group of Cr$_2$O$_3$ being $\underline{\overline{3}m}$ 
allows new 
tensors, i.e., polar c-tensors of odd rank and axial c-tensors of even 
rank. Thus,  electric 
dipole effects due to polar c-tensors of odd rank
 are allowed below $T_N$. 

In order to construct the nonlinear free energy for SHG, one needs to 
know the order parameter in Cr$_2$O$_3$. This can be achieved following 
Dzialoshinskii's \cite{dzi} derivation for the Landau theory 
of second order phase transitions \cite{tol}.
We assign a mean spin ${\bf S}_\beta$ ($\beta=1,2,3,4$)
to each of the four Cr$^{3+}$ ions. Out of the ${\bf S}_\beta$
one can form the  following linear combinations 
${\bf m }$, ${\bf l_1}$, ${\bf l_2}$ and ${\bf l_3}$:

\begin{eqnarray}  
{\bf m} = {\bf S_1} + {\bf S_2} + {\bf S_3} + {\bf S_4} \nonumber \\ 
{\bf l_1} = {\bf S_1} - {\bf S_2} - {\bf S_3} + {\bf S_4} \nonumber \\ 
{\bf l_2} = {\bf S_1} - {\bf S_2} + {\bf S_3} - {\bf S_4} \nonumber \\ 
{\bf l_3} = {\bf S_1} + {\bf S_2} - {\bf S_3} - {\bf S_4} 
\end{eqnarray}  
 The components of the vectors 
${\bf m}$ and ${\bf l_\alpha}$ ($\alpha=1,2,3$) can be
classified
according to the irreducible representations of 
the paramagnetic group $\overline{3}m$ as given in 
Table \ref{Table1}\cite{tol,tinkham}.
 The components $m_z$, $l_{1z}$, $l_{2z}$, $l_{3z}$ 
transform according to the one-dimensional representations $A_{2g}$, 
$A_{1g}$, $A_{1u}$, $A_{2u}$ of the point group $\overline{3}m$. The $x$ 
and $y$ components of the vectors ${\bf m}$ and ${\bf l_1}$ transform 
according to the two-dimensional representation $E_g$; and the $x$
 and $y$ components of the vectors  
${\bf l_2}$ and ${\bf l_3}$ transform according to $E_u$.
We make now the following key observation: 

{\sl The order
parameter of a magnet is given by that irreducible representation
of the paramagnetic point group which is invariant under
 the symmetries of the magnetic group.}

This statement can be considered as a generalization
 of von Neumann's principle (see \cite{bir}) and
 follows from the observation
that the thermodynamic expectation value of any combination
of the constituent spins ${\bf S}_\beta$ which is not
invariant under the magnetic group, i.e.\
$\underline{\bar3m}$ for the case of Cr$_2$O$_3$,
can be shown to vanish identically. For 
Cr$_2$O$_3$  the irreducible representation which is invariant
 under $\underline{\bar3m}$ is
$A_{1u}$, i.e.\ $l_{2z}$. 

Thus, the order parameter for Cr$_2$O$_3$,
 which is the staggered magnetization constructed 
from $l_{2z}$ is a c-axial scalar. The description given
by $l_{2z}$ corresponds to a collinear ordering (up,down,up,down)
which is the spin ordering in Cr$_2$O$_3$.  
Now, we can write down the free energy due to the 
electric dipole contribution as,  

\begin{eqnarray} 
F  = -\left[\,\chi_{ijk}(T>T_N)\> E_i^\ast \> E_j \> E_k \right.
\nonumber \\  \left.
+\ \chi^\ast_{ijk}(T>T_N)\> E_i \> E_j^\ast \> E_k^\ast
\,\right]\> l_{2z}
\label{f1}    
\end{eqnarray}  

\noindent and therefore the SHG susceptibility tensor can be obtained as,   

\begin{equation}  
\chi_{ijk}^{SHG}(T<T_N) =\chi_{ijk}(T>T_N)\> l_{2z}~.
\label{chishg}
\end{equation}  
i.e.\ a c-polar tensor of third rank is obtained from
 the tensorial product of an i-axial tensor of third rank with
 a c-axial scalar (or pseudoscalar). 
 Since the susceptibility tensor $\chi_{ijk}$ above $T_N$ 
is an i-axial tensor of third rank, we know all the non-zero components 
from the symmetry analysis \cite{bir}.  Thus, using 
Eq.\ (\ref{chishg}), one obtains all the non-zero components  
of the SHG $\chi$, which are, $\chi_{yyy}= -\chi_{yxx}= -\chi_{xyx}
=-\chi_{xxy}$. Furthermore,  
the symmetry of $\chi(T>T_N)$ in the paramagnetic phase dictates that 
$\chi^{SHG}(T<T_N)$ is linearly dependent on the order parameter, which  
is the reason why one observes the AFM domains through SHG in Cr$_2$O$_3$. 

The second example we want to illustrate is YMnO$_3$. YMnO$_3$ is
 a ferroelectromagnetic material
 whose crystal structure above the Curie 
temperature $T_c\approx 913 K$ is presumably centrosymmetric and
 described by the
 point group $6/mmm$. The elementary unit cell contains six
 Mn$^{3+}$ ions. 
Below $T_c$,  YMnO$_3$ orders ferroelectrically and the
 charge ordering breaks inversion symmetry. The YMnO$_3$ structure
 is then described by the 
non-centrosymmetric point group $6mm$. The vector
  ${\bf P}^{(SP)}=(0,0,{P_z}^{(SP)})$ 
of spontanous polarization
is directed along the six-fold z-axis. 
The magnetic properties of YMnO$_3$ arise from 
the manganese ions Mn$^{3+}$ in the high spin state $S=2$. 
Below the N\'eel temperature, $T_N\approx 74 K$, 
the spins of the six Mn$^{3+}$ ions in the unit cell are ordered 
antiferromagnetically in a triangular structure perpendicular to the 
polar axis. The crystallographic, magnetic and electric properties of 
the hexagonal YMnO$_3$ and the other rare-earth manganites were studied 
in the sixties and the related data are available in \cite{lan}. New data 
concerning dielectric, magnetic, infrared and Raman studies have been 
also reported recently \cite{hua,ili}.  

In the electric dipole approximation, SHG is allowed in YMnO$_3$ below 
$T_c$ due to the inversion symmetry breaking by
the ferroelectric ordering of charges \cite{fro}. Thus, from  
the symmetry of the susceptibility tensor in the paraelectric phase and  
that of the order parameter which is the spontaneous polarization   
$P_z^{(SP)}$ along the six-fold axis, one should be able
 to write down a free energy like Eq.\ 
(\ref{f1}). $6/mmm$ is centrosymmetric, which 
 implies that in the paraelectric phase only polar tensors of even rank
 and axial tensors of odd rank are allowed. $P_z^{(SP)}$ is an 
 i-polar first rank tensor (vector) directed along $z$. Therefore,
 in order to obtain an i-polar third rank tensor in the ferroelectric
 phase which should describe the SHG electric dipole contribution,
 only contractions of the $6/mmm$ symmetry polar tensors with odd
 powers of the order parameter are allowed.  In the lowest order,
 the free energy can be written as, 

\begin{eqnarray}  
F = -\left[\,\chi_{ijkz}(T>T_c) \> E_i^\ast \> E_j \> E_k    
\label{F_YMO_FE} \right. \\ \left.  
+\chi^\ast_{ijkz}(T>T_c) \> E_i \> E_j^\ast \> E_k^\ast
\,\right] \> P_z^{(SP)}\nonumber     
\end{eqnarray}  
such that the SHG susceptibility tensor in the
 ferroelectric phase is obtained as,  

\begin{equation}  
\chi_{ijk}^{SHG}(T_c>T>T_N) = \chi_{ijkz}(T>T_c) \> P_z^{(SP)}~.     
\label{chi_YMO_FE}
\end{equation}  
Here 
 the susceptibility tensor $\chi_{ijkl}$ above 
the Curie temperature $T_c$, i.e.\ in the paraelectric phase, is an 
i-polar tensor of fourth rank.   
Now, from the symmetry analysis \cite{bir} of this tensor,
 one gets the non-zero 
components of $\chi(T>T_c)$ in the paraelectric phase which when 
contracted with the order parameter   
give rise to all the non-zero components of $\chi^{SHG}$ in the
 ferroelectric phase, i.e., 
$\chi_{zzz}$ and $\chi_{xxz}(3) = \chi_{yyz}(3)$
 ((3) denotes all possible permutations of the 3 indices). 
Moreover, it follows from the above expression that the SHG 
susceptibility in the ferroelectric phase is a linear function of the 
ferroelectric order parameter.  

As already mentioned earlier, YMnO$_3$ belongs to the crystal 
class $6mm$ in the ferroelectric phase. Below $T_N$, 
 the six magnetic ions in the unit cell order 
 antiferromagnetically and perpendicular to the six-fold axis, i.e.\ 
three spins are arranged in a triangular structure on planes 
normal to the six-fold axis and separated from each other by a distance 
equal to half the lattice period along the hexagonal axis.   
The magnetic ordering in this material 
is non-collinear but coplanar and can be determined from the exchange 
interactions among the spins only, similar to that in Cr$_2$O$_3$. The
 corresponding magnetic group is $\underline{6}m\underline{m}$. 
Following
 Nedlin \cite{ned} and Pashkevich {\it {et al.}} \cite{pas}, 
we consider the following linear combinations of spins: 
    
\begin{eqnarray}  
{\bf s} = {\bf S_1} + {\bf S_2} + {\bf S_3} + {\bf S_4} + {\bf S_5} + 
{\bf S_6} \nonumber \\   
{\bf l} = {\bf S_1} + {\bf S_2} + {\bf S_3} - {\bf S_4} - {\bf S_5} - 
{\bf S_6} \nonumber \\  
{\bf \tau_1} = {\bf S_1} - \omega^{\ast} {\bf S_2} - \omega {\bf S_3} 
- {\bf S_4} + \omega^{\ast} {\bf S_5} + \omega {\bf S_6} \nonumber \\  
{\bf \tau_2} = - {\bf \tau_1}^{\ast} \nonumber \\  
{\bf \sigma_1} = {\bf S_1} - \omega^{\ast} {\bf S_2} - \omega {\bf S_3}
+ {\bf S_4} - \omega^{\ast} {\bf S_5} - \omega {\bf S_6} \nonumber \\
{\bf \sigma_2} = {\bf \sigma_1}^{\ast}~,
\label{sde}
\end{eqnarray}
where $\omega^{\ast}$ is the complex conjugate of the 
phase factor $\omega= e^{i\pi/3}$. The magnetic irreducible
 representations  
are given by some linear combinations $\psi$ of the spin components   
 (see Table \ref{Table2}). Since the spin 
ordering is coplanar, $\psi$ may easily 
be expressed in terms of the components of the vectors ${\bf s}$, 
${\bf l}$, ${\bf \tau}$ and ${\bf \sigma}$ written in the cyclic 
coordinate frame \cite{ned,pas} as follows. 

\begin{equation}  
{\bf s} = e^{-} s^{+} + e^{+} s^{-} + e_z s^z~,  
\end{equation}  
where $s^{\pm} = s^{x}\pm i s^{y}$ and similarly for  
 ${\bf l}$, ${\bf \tau}$ and ${\bf \sigma}$. Here the
${e_i}$'s are the unit vectors along the axes of the crystal coordinate 
frame (the z-axis coincides with the six-fold axis of the crystal lattice). 

 From Table \ref{Table2} we learn that $\psi_1, \cdots,\psi_4$ 
transform according to the one-dimensional representations $A_1$,
 $A_2$, $B_1$ and $B_2$ 
whereas $\psi_5$ and $\psi_6$ transform according
 to  the two-dimensional irreducible representations  $E_1$
 and $E_2$ of the 
paramagnetic and ferroelectric group $6mm$. The irreducible representation
 which remains invariant under all symmetry elements of the
 magnetic group $\underline{6}m\underline{m}$ is B$_1$, therefore
 $\psi_3$ is a good candidate to be defined as the antiferromagnetic
 order parameter \cite{pas} for YMnO$_3$.

$\psi_3$ is a complicated combination of the spin components of
 the six ions in the unit cell. Let us build a more intuitive
 object
 which belongs to the same $B_1$ irreducible
 representation and where not only spin components but also
 vector components are introduced, whose physical interpretation
 may be that of a local field on each Mn$^{3+}$ ion.
  Thus, similar to the spins ${\bf S_1} 
\cdots {\bf S_6}$, one can introduce i-polar vectors ${\bf V_1} 
\cdots {\bf V_6}$ (which might correspond to local planar displacements 
of the Mn$^{3+}$ ions) and form the linear combinations ${\bf p}$, 
${\bf q}$, ${\bf \eta}$ and ${\bf \rho}$ analogous to ${\bf s}$, ${\bf l}$, 
${\bf \tau}$ and ${\bf \sigma}$ in Eq.\ (\ref{sde}). 
These linear combinations of the i-polar 
vectors also follow the  irreducible representations of the spatial 
group $6mm$ .  From the direct product representations of the spin 
pseudovectors 
and that of the i-polar vectors (which of course is reducible), one  
obtains the following combination $\Lambda=\sigma_1^{+}\eta_2^{-} + 
\sigma_2^{-}\eta_1^{+}$ which becomes invariant under all the   
symmetry operations of the magnetic group.
Thus $\Lambda$ should be equivalent to $\psi_3$ in the sense
 that they belong to the same irreducible one dimensional representation.
 Furthermore,   
from the generating matrices of the magnetic group, one can also   
construct an invariant c-axial quantity in lowest order, which in the 
present case is a tensor of rank three, $R_3$ 
(in $\underline{6}m\underline{m}$, all the c-axial tensors of rank smaller 
than three 
vanish). Now expanding the 
free energy  in terms of $R_3$ one obtains in lowest order, 

\begin{eqnarray}  
F = -\left[\,\chi_{ijklmn}(T>T_N) \> E_i^\ast \> E_j \> E_k 
\nonumber \right.\\ \left.
+\ \chi^\ast_{ijklmn}(T>T_N) \> E_i \> E_j^\ast \> E_k^\ast 
\,\right] \> R_{lmn}    
\end{eqnarray}  
and thus the SHG susceptibility tensor can be obtained as, 

\begin{equation}  
\chi_{ijk}^{SHG}(T<T_N) = \chi_{ijklmn}(T>T_N)\> R_{lmn}
\label{SHG_YMO}  
\end{equation}  
It should be noted that the c-axial tensor $R_3$ has only 
one independent component, i.,e., $R_{yyy}=-R_{xxy}(3)$ which can
be identified as  
 the order parameter discussed above. Performing the
sum over $l$, $m$ and $n$ in Eq.\ (\ref{SHG_YMO})
explicitly we then obtain an alternative expression
for the allowed matrix elements contributing to
SHG in the antiferromagnetic phase as

\begin{eqnarray}
\chi_{ijk}^{SHG}(T<T_N)\ =\qquad\qquad \label{ED_YMO} \\ 
\left[\, \chi_{ijkyyy}(T>T_N)-\chi_{ijkxxy}(T>T_N)
\nonumber \right.\\ \left.
-\chi_{ijkxyx}(T>T_N) 
-\chi_{ijkyxx}(T>T_N)\, \right]\> \psi_3 \nonumber    
\end{eqnarray} 
Thus, from the symmetry properties of the sixth rank i-axial
tensor $\chi_{ijklmn}$
above $T_N$ \cite{fies}, together with Eq.\ (\ref{ED_YMO}), we get all the 
non-zero components of the 
SHG susceptibility tensor for $T<T_N$, which are, 
$\chi_{xxx}^{SHG}=-\chi_{yyx}^{SHG}
=-\chi_{yxy}^{SHG}=-\chi_{xyy}^{SHG}$
(as one can easily verify, see \cite{fies}).

Since YMnO$_3$ is characterized by two order parameters, one for the 
paraelectric-ferroelectric transition and the other for the 
paramagnetic-antiferromagnetic transition, it is natural to ask about 
the coupling between them. We observe here that
the SHG tensor in the antiferromagnetic phase, 
as given by Eq.\ (\ref{ED_YMO}), is directly proportional to
the even-rank i-axial tensor $\chi_{ijklmn}$
which is not allowed in
the high-temperature group $6/mmm$ for $T>T_c$, which is
centrosymmetric. $\chi_{ijklmn}$ need therefore to
be proportional to the ferroelectric order parameter
(compare Eq.\ (\ref{F_YMO_FE}) and Eq.\ (\ref{chi_YMO_FE})):

\begin{eqnarray}  
\chi_{ijklmn}(T>T_N)\ =\ \chi_{ijklmnz}(T>T_c)
\> P_z^{(SP)}~, \label{coupling} 
\end{eqnarray} 
where $\chi_{ijklmnz}$ is an i-axial tensor of rank
seven which is allowed in $6/mmm$. Comparing Eq.\ (\ref{coupling}) with
Eq.\ (\ref{ED_YMO}) we find immediately:

\begin{eqnarray}  
\chi_{ijk}^{SHG}(T<T_N) = \chi_{ijklmnz}(T>T_c) 
\> R_{l,m,n} \> P_z^{(SP)}  \ = \nonumber\\
\left[\, \chi_{ijkyyyz}(T>T_c)-\chi_{ijkxxyz}(T>T_c)
\qquad\qquad\nonumber \right.\\ \left.
-\chi_{ijkxyxz}(T>T_c) 
-\chi_{ijkyxxz}(T>T_c)\, \right]\> \psi_3\>P_z^{(SP)}~.
\nonumber
\end{eqnarray}  
It is therefore possible to derive
all the components of the SHG susceptibility tensor below $T_N$ from 
the non-zero components of $\chi_{ijklmno}$ together with both 
the order parameters. It is also clear from the above equation 
that the SHG susceptibility below $T_N$ is directly proportional 
to the bilinear combination of both  order parameters which   
in principle, could be verified from experiments.
 
The method described in the present manuscript to study the 
nonreciprocal optical effects in the magnetic materials through  
SHG is purely phenomenological and based on symmetry considerations. 
Therefore, it should be possible to generalize this phenomenology to 
other hexagonal rare-earth manganites RMnO$_3$ 
where R=Ho, Er, Tm, Yb, Lu etc.

In conclusion, we summarize the main findings of the present paper. 
A phenomenological Ginzburg-Landau theory has been developed for 
second harmonic generation in  materials which undergo
 one or more phase transitions
 by expanding the free energy in terms 
of the order parameter/s and the susceptibility tensor in the  
high-temperature  phase.  We have shown how to
obtain explicitly the SHG susceptibility
 components as a function of certain
susceptibility tensors allowed in the high-temperature phase
and of the order-parameter.
We have carried through this prescription for the
magnetoelectric compound Cr$_2$O$_3$ as well as for the
ferroelectric-antiferromagnetic material YMnO$_3$. We also argue 
that this analysis can be extended to the other hexagonal rare-earth 
manganites.   

\acknowledgements 
The authors would like to acknowledge several discussions with  
D. Fr\"ohlich, R. V. Pisarev, St. Leute, Th. Lottermoser, C. Reimpell  
and M. Fiebig. This work is supported by the Deutsche 
Forschungsgemeinschaft, the Graduiertenkolleg "Festk\"orperspektroskopie".

\end{multicols}


\newpage
\begin{table}
\[
\begin{array}{r|c|}  
&\multicolumn{1}{c|}{Irreducible \>\> Representations}\\
\hline 
A_{2g}\ &   
\begin{array}{c} {m_z}  
\end{array} \\
\hline
A_{1g}\ &
\begin{array}{c} {l_{1z}}     
\end{array} \\
\hline
A_{1u}\ &
\begin{array}{c} {l_{2z}}    
\end{array} \\
\hline
A_{2u}\ &
\begin{array}{c} {l_{3z}}  
\end{array} \\
\hline
E_g\ &
\left(\begin{array}{c} {m_x}, {m_y} \\ 
                       {l_{1x}}, {l_{1y}}    
\end{array}\right) \\
\hline
E_u\ &
\left(\begin{array}{c} {l_{2x}}, {l_{2y}} \\ 
                       {l_{3x}}, {l_{3y}}   
\end{array}\right) \\
\hline
\end{array}
\]
\caption{Classification of the components of the vectors $m$, $l$, 
in accordance with the irreducible representations 
of the crystallographic point group $\overline{3}m $.
\label{Table1}}   
\end{table}  
\begin{table}
\[
\begin{array}{r|c|c|}  
&\multicolumn{2}{c|}{Irreducible \>\> Representations}\\
\hline 
A_1\ &
\begin{array}{c} \psi_1 = -(\tau_1^{-}+\tau_2^{+})    
\end{array} \\
\hline
A_2\ &
\begin{array}{c} \psi_2 = (\tau_1^{-} - \tau_2^{+}) \\
                       \psi_2^\prime = s^z  
\end{array} \\
\hline
B_1\ &
\begin{array}{c}  \psi_3 = (\sigma_1^{-} + \sigma_2^{+}) \\
                       \psi_3^\prime = l^z 
\end{array} \\
\hline
B_2\ &
\begin{array}{c}  \psi_4 = (-\sigma_1^{-} + \sigma_2^{+}) 
\end{array} \\
\hline
E_1\ &
\begin{array}{c}  ({\psi_{5,I} = s^{+}, 
                        \psi_{5,II} = s^{-}}) \\ 
                        ({\psi_{5,III} = \sigma_2^{-}, 
                        \psi_{5,IV} = \sigma_1^{+}}) \\  
                        ({\psi_{5,V} = \tau_1^z, 
                        \psi_{5,VI} = -\tau_2^z})  
\end{array} \\
\hline
E_2\ &
\begin{array}{c}  ({\psi_{6,I} = l^{+}, 
                        \psi_{6,II} = l^{-}}) \\
                        ({\psi_{6,III} = -\tau_2^{-}, 
                        \psi_{6,IV} = \tau_1^{+}}) \\
                        ({\psi_{6,V} = \sigma_1^z, 
                        \psi_{6,VI} = -\sigma_2^z})  
\end{array} \\
\hline
\end{array}
\]
\caption{Classification of the components of the vectors $s$, $l$, 
$\tau$ and $\sigma$ in accordance with the irreducible representations 
of the crystallographic point group $6mm$.\label{Table2} }   
\end{table}  
\end{document}